\documentclass[aps,preprint,floatfix,nofootinbib,showpacs]{revtex4-1}
\pdfoutput=1
\usepackage{graphicx,color}
\usepackage{hyperref}

\def\lsim{\:\raisebox{-0.5ex}{$\stackrel{\textstyle<}{\sim}$}\:}
\def\gsim{\:\raisebox{-0.5ex}{$\stackrel{\textstyle>}{\sim}$}\:}

\begin{document}

{\small
\begin{flushright}
IUEP-HEP-19-01
\end{flushright} }

\title{Vector-like Quark Interpretation of Excess in Higgs Signal Strength}

\renewcommand{\thefootnote}{\arabic{footnote}}

\author{
Kingman Cheung$^{1,2,3,4}$, Wai-Yee Keung$^{5,1}$, Jae Sik Lee$^{6,7}$, and
Po-Yan Tseng$^{8,1}$}
\affiliation{
$^1$ Physics Division, National Center for Theoretical Sciences,
Hsinchu, Taiwan \\
$^2$ Department of Physics, National Tsing Hua University,
Hsinchu 300, Taiwan \\
$^3$ Division of Quantum Phases and Devices, School of Physics, 
Konkuk University, Seoul 143-701, Republic of Korea \\
$^4$ Department of Physics, National Central University, Chungli, Taiwan\\
$^5$ Department of Physics, University of Illinois at Chicago,
Illinois 60607 USA \\
$^6$ Department of Physics 
Chonnam National University, \\
300 Yongbong-dong, Buk-gu, Gwangju, 500-757, Republic of Korea \\
$^7$ Institute for Universe and Elementary Particles,
Chonnam National University, \\
300 Yongbong-dong, Buk-gu, Gwangju, 500-757, Republic of Korea\\
$^8$Kavli IPMU (WPI), UTIAS, The University of Tokyo, 
Kashiwa, Chiba 277-8583, Japan
}
\date{\today}

\begin{abstract}
  There is a $+2\sigma$ deviation in the average Higgs-signal strength for
  all the $7+8+13$ TeV data up to Summer 2018. We find that a slight
  reduction of the bottom-Yukawa coupling can fit the data better than
  the standard model.  We suggest an extension with a  vector-like quark
  doublet, of which the right-handed component of $b'$ mixes nonnegligibly
  with the standard model $b$ quark. We show that the mixing would induce
  a reduction of the bottom Yukawa coupling. 
  Simultaneously,
  the coupling of the $Z$ boson to the right-handed $b$
  quark increases, which could reduce the forward-backward asymmetry of
  bottom production at LEP and bring it closer to the experimental value.
\end{abstract}

\maketitle

\section{Introduction}

The standard model (SM) like Higgs boson was discovered in 2012
\cite{Aad:2012tfa,Chatrchyan:2012xdj}.
After Run I at 7 and 8 TeV, the identity of the Higgs boson was very close
to the SM one \cite{Cheung:2013kla,Cheung:2014noa}.
With more and more measurements of the Higgs-signal strengths for various
production and decay channels at a center-of-mass energy of 13 TeV,
including the newly established $t\bar tH$
production channel \cite{Sirunyan:2018shy,Aaboud:2018urx},
and $H\to b\bar b$ \cite{Sirunyan:2018kst,Aaboud:2017xsd}
and $\tau\tau$ \cite{Sirunyan:2017khh,ATLAS:2018lur} in 2018, the SM-like
Higgs boson is further confirmed. The most updated fits to the Higgs-boson
couplings in various scenarios have recently been performed \cite{Cheung:2018ave}.

Some very interesting results emerge from the new global fits, which 
were not realized previously.  The combined average signal strength of
the Higgs boson now stands at a $+2\sigma$ deviation from the SM value,
namely $\mu_{\rm exp} = 1.10 \pm 0.05$.  Note that from the earlier
combined signal strength at $7+8$ TeV, the ATLAS and CMS obtained
\cite{Khachatryan:2016vau}
\[
\mu_{7+8\,{\rm TeV}} = 1.09 ^{+0.11}_{-0.10} \;,
\]
which was about $1\sigma$ above the SM. The 13 TeV data continues to show
such trend, by combining all production and decay channels
\cite{Cheung:2018ave}:
\[
\mu_{13,{\rm TeV}} = 1.10 \pm 0.06\,.
\]
These two results can be combined naively into a total signal strength
\begin{equation}
\mu_{\rm All} = 1.10 \pm 0.05 \;,
\end{equation}
which shows a $2\sigma$ above the SM prediction.

If the overall signal strength continues to about 10\% above the
SM prediction while the uncertainties continues to reduce, it would
pose a threat to the SM Higgs boson. One of the most economical fits
to the Higgs-signal strength is to vary the total width of the Higgs boson.
In Ref.~\cite{Cheung:2018ave} we found that the best-fit value for
the $\Delta \Gamma_{\rm tot}$ is
\begin{equation}
\Delta \Gamma_{\rm tot} = -0.285 ^{+0.18}_{-0.17} \; {\rm MeV}
\end{equation}
which means a reduction of about $0.285 / 4 = 0.07$
or 7\% to the total width.

Naively, it is hard to imagine that one can add any new channels
to {\it reduce} the total width. Nevertheless, one possibility is to
reduce the partial width into $b\bar b$ with a reduced bottom-Yukawa coupling,
provided that the current uncertainty of $H\to b \bar b$
coupling is of order 20\%.
There are a few obvious possibilities that one can consider:
(i) a $b'_{L/R}$ singlet vector-like quark model but the left-handed (LH)
component would modify the
CKM phenomenology significantly, and thus subject to severe constraints.
(ii) A $(t',b')_{L/R}$ vector-like quark doublet with
hypercharge $Y/2 = 1/6$ (the same as the SM quark doublet) but it would
increase the tension with the experimental bottom forward-backward asymmetry
at $Z$-pole.
In this work, we explore an extension\cite{Chang:1996pf} of the SM by adding an $SU(2)$
vector-like quark doublet with a different
hypercharge of $\frac{Y}{2} = -\frac{5}{6}$,
of which the upper component $b'^{-\frac{1}{3}}_R$ mixes with the SM
$b_R$ quark while $b_L'^{-\frac{1}{3}}$ mixes negligibly with $b_L$. In such a
way, the right-handed (RH) component of the bottom quark is reduced and thus
the bottom-Yukawa coupling is reduced with respect to the SM value.
Therefore, it can effectively explain why the average Higgs-signal strength
is enhanced.

Historically, the measurement of forward-backward asymmetry ${\cal A}_{FB}^b$
of the bottom quark at the $Z^0$ pole
remains a $-2.4\sigma$ deviation from the SM prediction
\cite{Tanabashi:2018oca}. In the present context, due to the mixing between  
$b'^{-\frac{1}{3}}_R$ and $b_R$ the effective RH coupling of bottom quark to
the $Z$ boson is enhanced, such that the ${\cal A}_{FB}^b$ would decrease\cite{Chang:1998uj}
in accord with the experimental data.
Apparently, the addition of the new vector-like quark doublet can
simultaneously explain
the Higgs-signal strength $\mu_{\rm Higgs}$ and the forward-backward
asymmetry ${\cal A}_{FB}^b$ in the correspondingly right direction.
However, there are other precision constraints that we have to consider,
namely, the ratio $R_b$ of the partial width $Z\to b \bar b$ to the
total hadronic width, as well as the total hadronic width of the $Z$ boson.
We will give details in subsequent sections.

The organization of the paper is as follows. In the next section, we
describe the extension of an isospin doublet of vector-like quarks, and
modifications to the Higgs and $Z$ couplings. In Sec. III, we fit
the parameter $\delta \equiv \Delta / M$,
  where $\Delta$ measures the mixing and $M$ is approximately the mass of
  heavy vector-like quark,
to the data of Higgs-signal strengths and to
${\cal A}_{FB}^b$, with or without $R_b$ and $\Gamma_{\rm had}$.
We discuss some other potential issues with
modifications of the bottom-quark couplings and then conclude.

\section{Formalism}

  In this work, we consider the vector-like quark doublet
  with a hypercharge $\frac{Y}{2} = -\frac{5}{6}$ denoted by
\[ {\cal B}_{L,R}=
\left(\begin{array}{c} b'^{-\frac{1}{3}} \\
   p'^{-\frac{4}{3}} \end{array}\right)_{L,R}
\ ,\quad
\left(\frac{Y}{2}\right)_{\cal B}=-\frac{5}{6} \ .
\]
We label electric charges of the new particles $b',p'$ by superscripts. 
Since $b'$ carries the same electric charge as the SM
bottom quark $b$, they can mix.
The quark mass matrix of  $(b,b')$ receives additional contributions from the 
following new coupling with the SM Higgs doublet $H$,
\begin{equation}
{\cal L} \supset 
g_{\cal B} \overline{ {\cal B}_{L} }\widetilde H b_R + \hbox{ h.c. }
=
g_{\cal B} (\overline{b'_{L}} \ ,\  \overline{p'_L} \ )
\left(\begin{array}{c}
-\frac{1}{\sqrt2}(v+h)  \\
H^- \end{array} \right) b_R   + \hbox{ h.c. }\;,
\end{equation}
where $\widetilde{H} = i \tau_2 H^*$.
Note that the vector-like quarks receive their mass $M$ from some mechanisms,
other than the usual electroweak symmetry breaking. We assume $M$ is of order
TeV or more.
Then the quark-mass matrix and the interactions with the SM Higgs boson
become
\begin{equation}
  \label{yukawa}
{\cal L}_{Y} \supset - (\overline{b_L} \ ,
\ \overline{b'_L} \ ) \left( \begin{array}{cc} m(1+\frac{h}{v}) & 0
  \\ \frac{g_{\cal B} v}{\sqrt 2}(1+\frac{h}{v} ) & M \end{array} \right)
\left( \begin{array}{c}   b_R \\
  b'_R \end{array} \right ) \;
+ \; 
\hbox{ h.c.}
\end{equation}
The large mass $M$ for the vectorial $\cal B$ is unrelated to $H$.
It is much larger  than the off-diagonal mass 
$\frac{g_{\cal B} v}{\sqrt 2} \equiv \Delta$.
The mass parameter $m$ accounts for the $b$-quark mass in the SM 
if we ignore $\Delta$. 

\subsection{Mass diagonalization and Modifications to Bottom Yukawa}

{}From the above equation, the mass terms for $b,b'$ can be written as
\begin{equation}
  \label{quark-mass}
        {\cal L}_{\rm mass} \supset - (\overline{b_L} \ ,
\ \overline{b'_L} \ ) \left( \begin{array}{cc} m  & 0 \\
   \Delta  & M \end{array} \right)
\left( \begin{array}{c}   b_R \\
  b'_R \end{array} \right ) \;
+ \; 
\hbox{ h.c.}
= - (\overline{b_L} \ ,
\ \overline{b'_L} \ ) {\cal M}
\left( \begin{array}{c}   b_R \\
  b'_R \end{array} \right ) \;
+ \; 
\hbox{ h.c.}
\end{equation}
where $\Delta = g_{\cal B} v / \sqrt{2} $.

We require the following
left and right rotations to diagonalize the non-hermitian 
mass matrix:
\begin{equation}
  \left ( \begin{array}{c}
    b \\
    b' \end{array} \right )_{L.R} =
\left ( \begin{array}{cc}
    \cos\theta_{L,R} &  \sin \theta_{L,R} \\
    - \sin\theta_{L,R} & \cos\theta_{L,R} \end{array} \right ) \;
  \left( \begin{array}{c}
    b \\
    b'
  \end{array} \right )_{L.R}^{m}
  \equiv
      {\cal U}_{L,R} \;
        \left( \begin{array}{c}
    b \\
    b'
        \end{array} \right )_{L.R}^{m}
\end{equation}
where the superscript ``$m$'' denotes the mass eigenstates.
For convenience we will drop the ``$m$'' wherever it is understood
to be the mass eigenstates.
The presence of the zero entry in the upper-right corner
of the quark-mass matrix in Eq.~(\ref{quark-mass})
suggests that the right rotation angle $\theta_R$
is of order $\frac{\Delta}{M}$, which is much larger than
the left rotation angle $\theta_L$ of order
$\frac{m\Delta}{M^2}$ for the favorable scenario
$ \Delta \gg m$. The suppression of $\theta_L/\theta_R$
is of order $m_b / O({\rm TeV}) \sim 10^{-3}$. 

More precisely, the non-hermitian mass matrix ${\cal M}$ is diagonalized by 
a bi-unitary rotation as
  \begin{equation}
    {\cal U}^\dagger_L \, {\cal M} \, {\cal U}_R = {\cal M}_{\rm diag} \;,
  \end{equation}
  which can be turned into
  \begin{equation}
    {\cal U}^\dagger_L \, {\cal M} {\cal M}^\dagger \, {\cal U}_L =
    {\cal U}^\dagger_R \, {\cal M}^\dagger {\cal M} \, {\cal U}_R =
    {\cal M}^2_{\rm diag} =
    \left( \begin{array}{cc}
      m_1^2 & 0 \\ 0 &  m_2^2
    \end{array}\right )  \;,
  \end{equation}
  with $m_1<m_2$. Then the hermitian mass matrix squared can be diagonalized
  and the corresponding eigenvalues and eigenvectors can be calculated
  exactly:
  \begin{eqnarray}
m_{1,2}^2 &=& \frac{(m^2+\Delta^2+M^2)\mp\sqrt{(m^2+\Delta^2+M^2)^2-4m^2M^2}}{2}
\nonumber \\
\sin2\theta_L &=&\frac{2m\Delta}{\sqrt{(m^2+\Delta^2+M^2)^2-4m^2M^2}} \\
\sin2\theta_R &=&\frac{2\Delta M}{\sqrt{(m^2+\Delta^2+M^2)^2-4m^2M^2}}
\end{eqnarray}
  In the limit $M,\Delta \gg m$, they can be simplified to
  \begin{equation}
  m_{1}^2 = \frac{m^2}{1 + \frac{\Delta^2}{M^2}}, \qquad
  m_{2}^2 = \Delta^2 + M^2 \;.
  \end{equation}
  The mixing angles can also be simplified as
  \begin{equation}
    \sin\theta_L \equiv s_L  \simeq \frac{m \Delta}{M^2 + \Delta^2}\,\qquad
    \cos\theta_L \equiv c_L \simeq 1 - \frac{1}{2}
\left ( \frac{m\Delta}{M^2 + \Delta^2} \right)^2  \;,
  \end{equation}
  and
\begin{eqnarray}
\label{eq:csR}
%
\sin\theta_R\equiv s_R \simeq \frac{\Delta}{\sqrt{M^2+\Delta^2}}\,,
\qquad
\cos\theta_R\equiv c_R \simeq \frac{M}{\sqrt{M^2+\Delta^2}}\,.
\end{eqnarray}
  In the above, we identify $m_1 = m_b$, the mass of the observed $b$-quark
  mass, and $m_2 = M_{b'}$ to be the TeV mass of the heavy vector-like quark.
  Practically, we can take $c_L \simeq 1$ in the analysis
and then we find the $h$-$b^m$-$b^m$ Yukawa coupling
depends only on one parameter of $\delta\equiv \Delta / M$.
%
More precisely,
the coupling for $(h/v) \overline{b^m_L} b^m_R$ is given by
\begin{equation}
  m c_L c_R - \Delta s_L c_R \simeq \frac{m}{1 + \delta^2} c_R
  \simeq m_b\frac{1}{\sqrt{1+\delta^2}}  c_R 
\end{equation}
where we use $\Delta  s_L = m {\delta^2}/{(1 + \delta^2)}$
and $c_R$ is given by Eq.~(\ref{eq:csR}) in the $m\to 0$ limit.
The result is an overall reduction in the Higgs Yukawa coupling by
$C_b\equiv c_R/\sqrt{1+\delta^2} $  from the SM value. 

There are also couplings for other off-diagonal elements, as given
in this equation:
\begin{equation}
{\cal L}_{Y} \supset - \frac{h}{v}\, (\overline{b^m_L} \ ,
\ \overline{b^{'m}_L} \ ) \left( \begin{array}{cc}
 m_b (1 + \delta^2)^{-1/2} c_R   &  
 m_b (1 + \delta^2)^{-1/2} s_R 
\\
  \Delta c_R  &   \Delta s_R  \end{array} \right)
\left( \begin{array}{c}   b^m_R \\
  b^{'m}_R \end{array} \right ) \;
+ \; H.c.
\end{equation}
We can immediately see that the off-diagonal
coupling of $h \overline{b'^m_L} b^m_R $
will dominate over the other one.  Phenomenologically, the so-produced
$b'$ will decay into $ h + b_R$. We shall discuss the collider signature
in the Discussion.

In the following we can focus on the effect of RH mixing
in numerical analysis.

\subsection{Modifications to the $Z$ couplings}

  In the weak eigenbasis, according to $T_{3f} - Q_f x_w$,   
the  $Z$ couplings to fermions $b_{L,R}$ and  $b'_{L,R}$,
are given by
\[
  - {\cal L}\supset
  g_Z (\overline{b}_L \ ,\ \overline{b'}_L \ ) \gamma^\mu Z_\mu 
  \left( \begin{array}{cc}
      - \frac{1}{2} + \frac{1}{3} x_w & 0\\
       0  & \frac{1}{2} +\frac{1}{3} x_w  \end{array} \right) 
\left(\begin{array}{cc} b_L \\ b'_L\end{array}\right) 
\]
\begin{equation}
 +  g_Z (\overline{b}_R \ ,\ \overline{b'}_R \ ) \gamma^\mu Z_\mu 
\left( \begin{array}{cc} \frac{1}{3} x_w & 0
  \\  0  & \frac{1}{2} +\frac{1}{3} x_w  \end{array} \right) 
\left(\begin{array}{cc} b_R \\ b'_R\end{array}\right) \,.
\end{equation}
After rotating into mass eigenbasis we have
\[
  - {\cal L}\supset
  g_Z (\overline{b^m_L} \ ,\ \overline{b^{'m}_L} \ ) \gamma^\mu Z_\mu 
  \left( \begin{array}{cc}
      - \frac{1}{2}(c_L^2 -s_L^2)  + \frac{1}{3} x_w &  - c_L s_L\\
      -c_L s_L  & \frac{1}{2}(c_L^2 - s_L^2) +\frac{1}{3} x_w
   \end{array} \right) 
  \left(\begin{array}{cc} b^m_L \\ b^{'m}_L\end{array}\right)
\]
\begin{equation}
 + g_Z (\overline{b^m_R} \ ,\ \overline{b^{'m}_R} \ ) \gamma^\mu Z_\mu 
 \left( \begin{array}{cc}
    \frac{1}{2} s_R^2 + \frac{1}{3} x_w & -\frac{1}{2} c_R s_R \\
    - \frac{1}{2} c_R s_R  & \frac{1}{2} c_R^2 +\frac{1}{3} x_w
 \end{array} \right) 
\left(\begin{array}{cc} b^m_R \\ b^{'m}_R\end{array}\right) \,.
\end{equation}
Here the gauge coupling $g_Z=g_2/\cos\theta_w$ and the electroweak
mixing $x_w=\sin^2\theta_w$.
Note that the $Z$ coupling to the LH $b$ quark is practically
the same as the SM coupling for a very small $s_L\approx \frac{m\Delta}{M^2}$.
On the other hand, the RH $b$ quark coupling is modified by an amount
$s_R^2/2 \sim \Delta^2 / (2 M^2)$.

There are a number of observables that would be modified when
the RH coupling to the $Z$ boson is modified:
\begin{enumerate}
\item {\bf Total hadronic width}.
  At tree level, the change to the decay width into $b\bar b$ is given by
  \begin{equation}
    \delta \Gamma_b^{\rm BSM} = \left[\Gamma^{\rm BSM,b}_{\rm tree}
      - \Gamma_{\rm tree}^{\rm SM,b } \right ]\,
     \left ( 1+ \frac{\alpha_s(M_Z)}{\pi} \right ) \;.
  \end{equation}
  With this modification the total hadronic width is changed to
  \begin{equation}
    \Gamma^{\rm BSM}_{\rm had} = \Gamma_{\rm had}^{\rm SM} +
    \delta \Gamma^{\rm BSM}_b \;.
  \end{equation}

\item $\mathbf{R_b}$.  The $R_b$ is the fraction of hadronic width into
  $b\bar b$, and so it is given by
  \begin{equation}
    R_b = \frac{ \Gamma_b^{\rm SM} + \delta \Gamma^{\rm BSM}_b}
    {\Gamma_{\rm had}^{\rm SM} + \delta \Gamma^{\rm BSM}_b} \;.
   \end{equation}
  The value of $R_b$ can increase for a moderate $s_R$ 
when $s_R\gg s_L$.

\item $\mathbf{ {\cal A}^b_{\rm FB}}$.
There is a large tension in the forward-backward asymmetry of $b$ quark
production at the $Z$ resonance,
\begin{equation}
{\cal A}^b_{\rm FB}
=\frac{3}{4} \times
             \frac{(g^e)_L^2-(g^e)_R^2}{(g^e)_L^2+(g^e)_R^2} \times
             \frac{(g^b)_L^2-(g^b)_R^2}{(g^b)_L^2+(g^b)_R^2} \ . 
\end{equation}
Those couplings to the $Z$ boson are basically given by 
$T_3-Qx_w$ in SM. For the electron it is simply
\[
\frac{(g^e)_L^2-(g^e)_R^2}{(g^e)_L^2+(g^e)_R^2}
= 
\frac{(-\frac{1}{2}+x_w)^2-x_w^2}{(-\frac{1}{2}+x_w)^2+x_w^2}
\]
while for the $b$ quark it is
\[
\frac{(g^b)_L^2-(g^b)_R^2}{(g^b)_L^2+(g^b)_R^2}=
\frac{(-\frac{1}{2}+\frac{1}{3}x_w)^2-\frac{1}{9}x_w^2}{(-\frac{1}{2}+
  \frac{1}{3} x_w)^2+\frac{1}{9} x_w^2} \;.
\]
Correspondingly, the modified forward-backward asymmetry is given by
\begin{equation}
  {\cal A}_{FB}^b = \frac{3}{4} \times
  \frac{(-\frac{1}{2}+x_w)^2-x_w^2}{(-\frac{1}{2}+x_w)^2+x_w^2}
  \times
 \frac{(-\frac{1}{2}(c_L^2-s_L^2) +\frac{1}{3} x_w)^2
    -(\frac{1}{2} s_R^2+\frac{1}{3}x_w)^2}
     {(-\frac{1}{2}(c_L^2-s_L^2) +\frac{1}{3} x_w)^2
       +(\frac{1}{2} s_R^2+\frac{1}{3}x_w)^2} \ .
\end{equation}

\end{enumerate}

\section{Fitting to data}
Four data sets are considered in our analysis.
They are summarized in the following table.
\begin{center}
\begin{tabular}{lll}
\hline  \hline
 Experimental Data &  SM values  & $\chi^2 ({\rm SM})$ \\
  \hline
  Higgs-signal strengths with the average  \hspace{0.2in} & & \\
   \hspace{0.2in}  $\mu_{\rm Higgs} =1.10 \pm 0.05$ & $\mu^{\rm SM} = 1.00$ 
   \hspace{0.2in} & $53.81$   \cite{Cheung:2018ave}  \\
  $\left( {\cal A}^b_{\rm FB} \right)^{\rm EXP}=0.0992\pm 0.0016$ &
   $0.1030 \pm 0.0002$ & $ 5.29$ \cite{Tanabashi:2018oca} \\
  $ R^{\rm EXP}_b=0.21629\pm 0.00066$ & $0.21582 \pm 0.00002$ & 
    $0.49$   \cite{Tanabashi:2018oca} \\
  $\Gamma_{\rm had}=1.7444\pm 0.0020 \;{\rm GeV}$ &
  $ 1.7411\pm 0.0008$ & $2.35$  \cite{Tanabashi:2018oca}\\
  \hline
\end{tabular}
\end{center}

The 125 GeV Higgs-signal strengths include a combined ATLAS+CMS
analysis for the 7+8 TeV datasets \cite{Khachatryan:2016vau} and
all the most updated 13 TeV data summarized in Ref.~\cite{Cheung:2018ave}.
The average signal strength is $\mu_{\rm Higgs} =1.10 \pm 0.05$
\cite{Cheung:2018ave}.
There are totally 64 data points. The goodness of the SM description for the
Higgs data stands at $\chi^2/d.o.f. = 53.81/64$, which gives a $p$-value of
$0.814$. As explained in Introduction, a reduction in the total Higgs
decay width can provide a better description of the Higgs data with
$\chi^2/d.o.f. = 51.44/63$, corresponding to a $p$-value of 0.851
\cite{Cheung:2018ave}. In this work, the reduction in the total width
is achieved by a slight reduction in the RH bottom Yukawa coupling.
On the other hand, the other three datasets were from the LEPI
precision measurements tabulated in PDG \cite{Tanabashi:2018oca}.
There has been a $2.4\sigma$ deviation in the ${\cal A}_{FB}^b$ while $R_b$
is very much consistent with the SM.

\begin{figure}[t!]
\centering
\includegraphics[height=3.2in,angle=270]{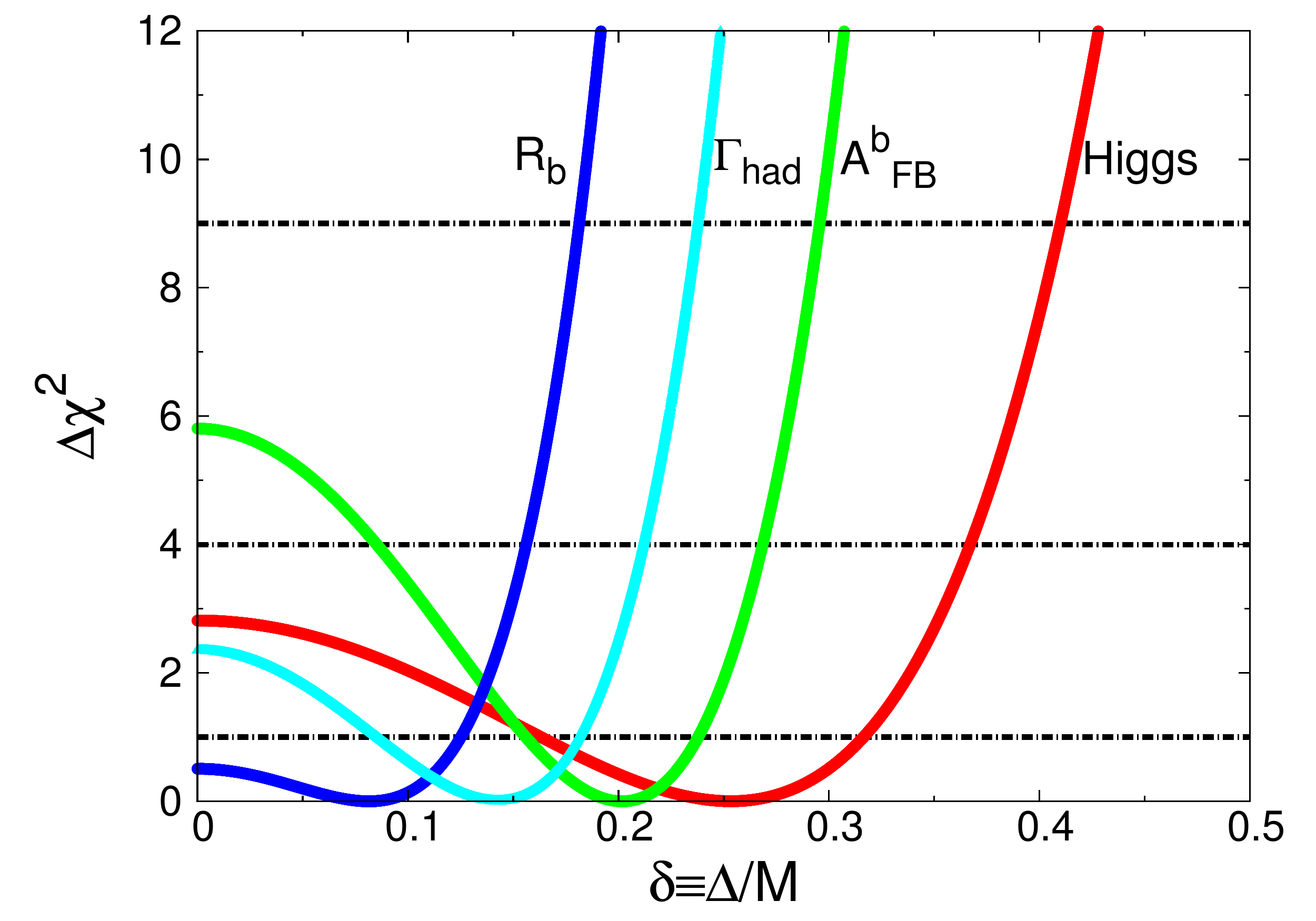}
\includegraphics[height=3.2in,angle=270]{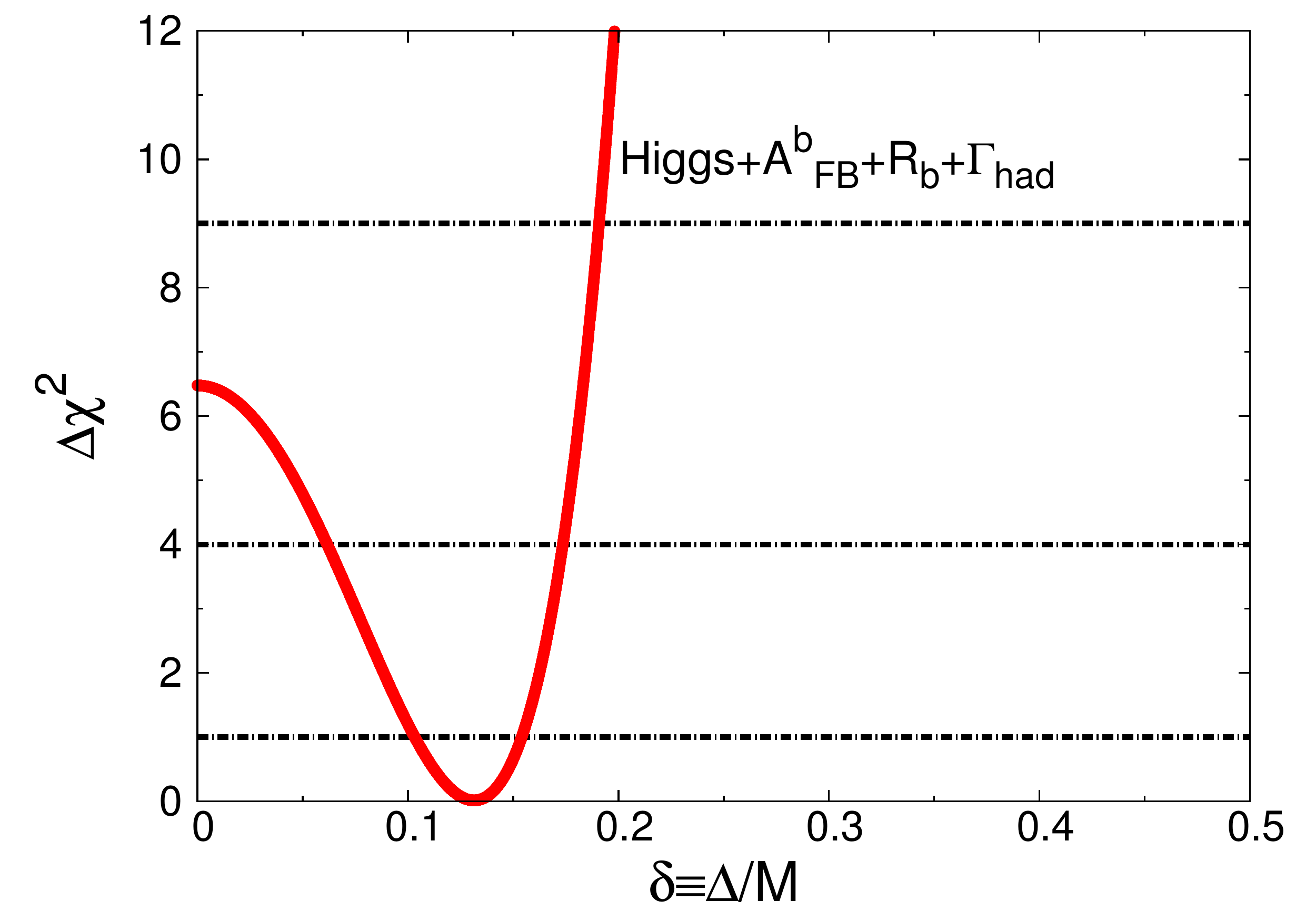}
\includegraphics[height=3.2in,angle=270]{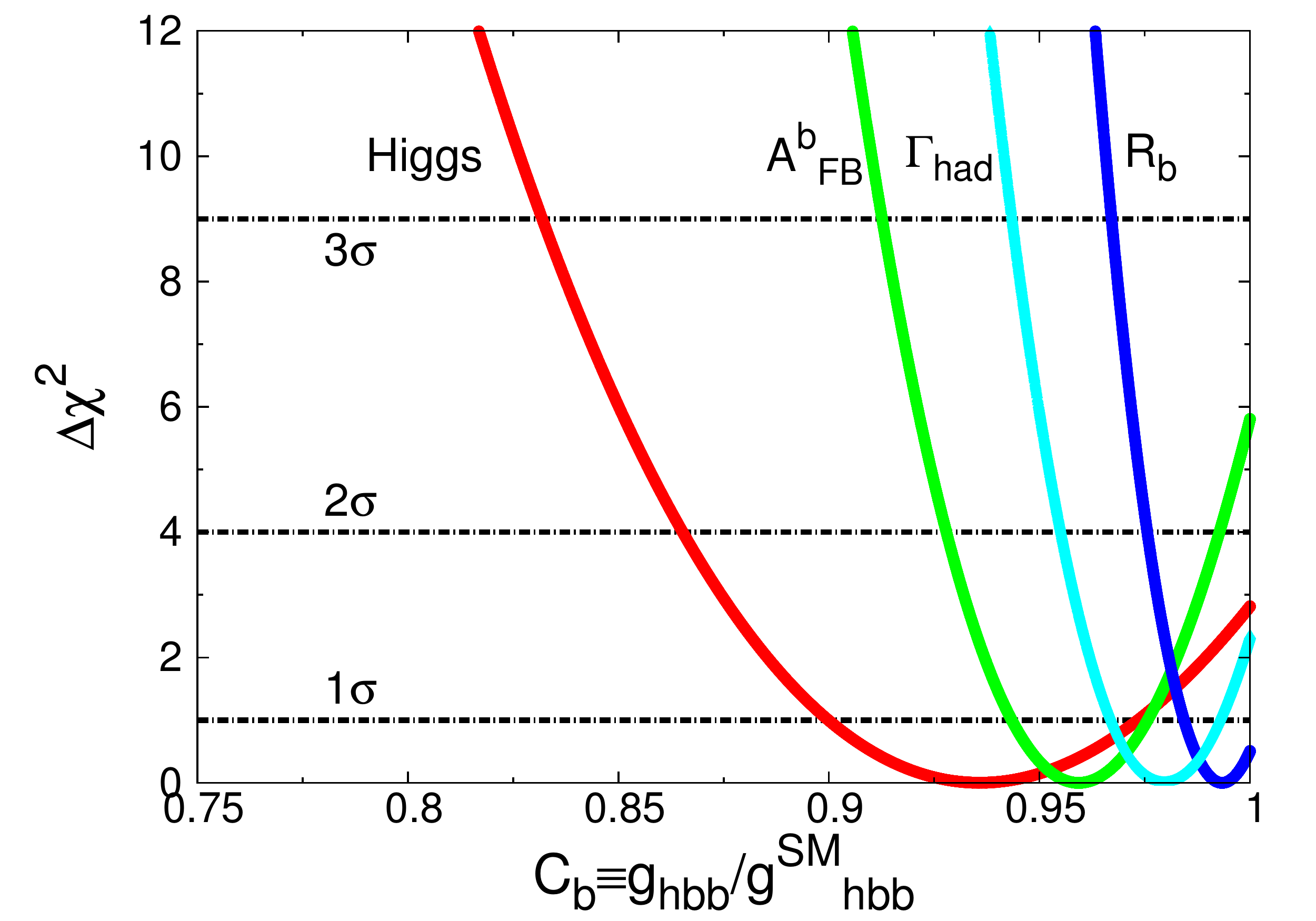}
\includegraphics[height=3.2in,angle=270]{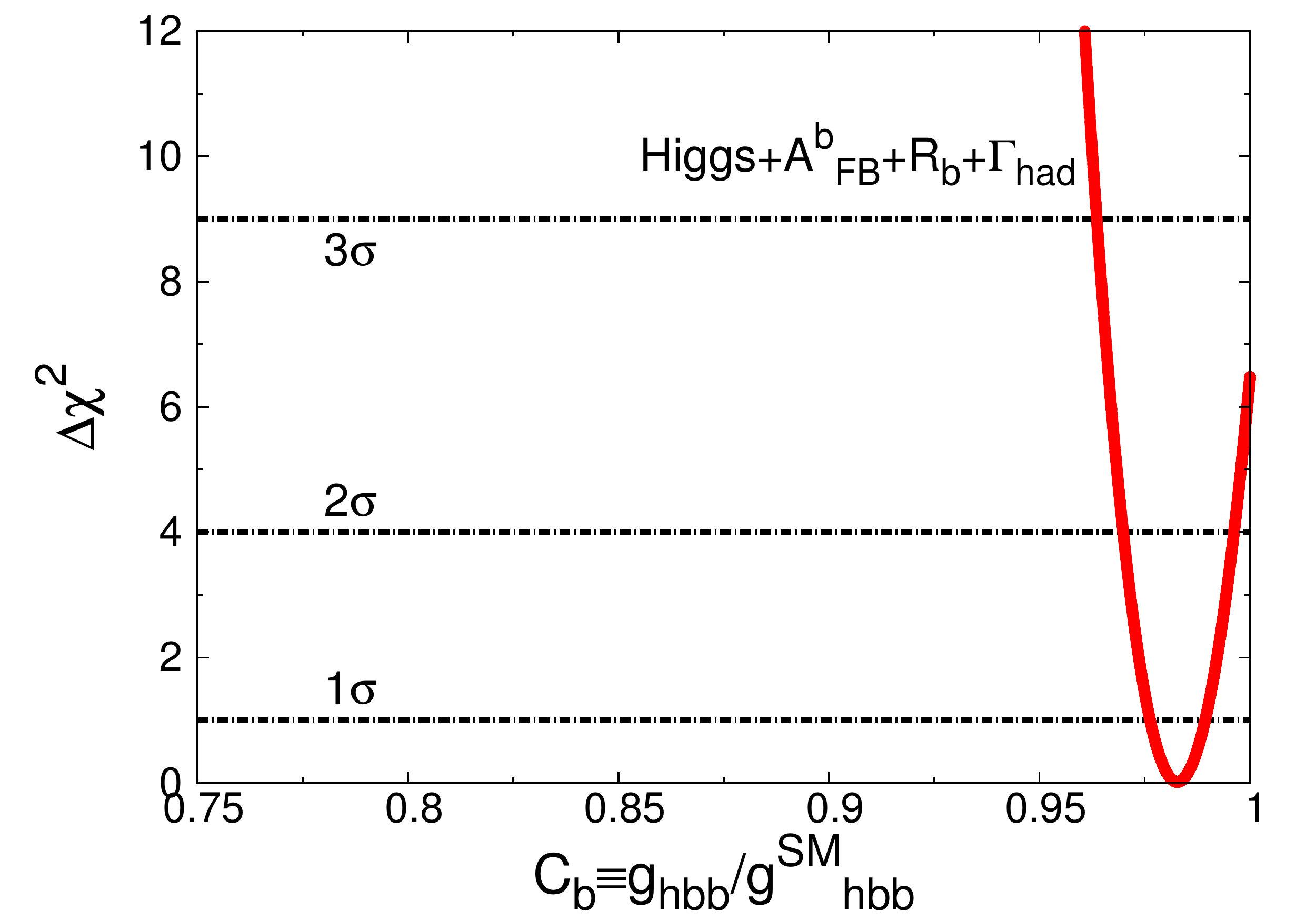}
\caption{\small \label{fig-1}
  Case-{\bf i} to {\bf v}: $\delta$ is varied while taking
  $c_L=1$ and $x_w=0.23154$.
Left-column: $\Delta \chi^2$ distributions
versus $\delta \equiv \Delta /M$ and versus
  $C_b \equiv g_{h b b} / g_{h b b}^{\rm SM}$ for individual fitting to
four experimental datasets:
(i)\,Higgs-signal strength,
(ii)\,$({\cal A}^b_{\rm FB})^{\rm EXP}$,
(iii)\,$R^{\rm EXP}_b$, and
(iv)\,$\Gamma_{\rm had}$, which correspond to
Case-{\bf i} to {\bf iv} in TABLE~\ref{tab-2}.
Right-column:$\Delta \chi^2$ distributions
versus $\delta \equiv \Delta /M$ and versus $C_b$
for the combined fitting,  
which corresponds to Case-{\bf v} in TABLE~\ref{tab-2}.
}
\end{figure}

In the following, we present our numerical results on fitting to
different combinations of the datasets with variation in 
$\delta\equiv \Delta/M$ and
a fixed or varying $x_w$.
We first show the fits with each single dataset listed in
the previous table.
Figure~\ref{fig-1} shows the $\Delta \chi^2$ distribution versus $\delta$
fitting to four single experimental datasets with 
a fixed $x_w=0.23154$ \cite{Tanabashi:2018oca} and $c^2_L=1$.
(Note that in the mass hierarchy $m \ll \Delta \ll M$ that we have assumed,
$c^2_L$ is practically equal to 1.)
The best fit values and uncertainties of $\delta $ for each dataset are
listed in Table~\ref{tab-2} from Case-{\bf i} to {\bf iv}.
We can see that the dataset on Higgs-signal strengths and that on 
$\left({\cal A}^b_{\rm FB}\right)^{\rm EXP}$ prefer a sizable mixing between
$b_R$ and $b'_R$, corresponding to the mixing angle equal to
$s_R \simeq \delta  \simeq 0.25$ and $0.20$, respectively.
The total hadronic width
mildly prefers a mixing with mixing angle equal to $s_R \simeq \delta \simeq
0.14$.
However, the $R_b$ is very much consistent with the SM and 
indicates a very small mixing $s_R \simeq \delta \simeq 0.08$
between $b_R$ and $b'_R$. 

Next we come to various combinations of datasets.
In Case-{\bf v}, we perform the fit by combining all
four experimental datasets by varying 
$\delta\equiv \Delta/M$ with a fixed $x_w = 0.23154$.
The result is shown in right-panels of Fig.\ref{fig-1} and Table~\ref{tab-2}.
The central value of $\delta $ shifts slightly to $\delta \simeq
  0.13$, which gives mild improvements to all four datasets.
Overall, the $\chi^2$ improves considerably.

  \begin{table}[thb!]
\caption{\small \label{tab-2}
  The best fit points to the experimental datasets: Higgs-signal strengths,
${\cal A}^b_{FB}$, $R_b$, and $\Gamma_{\rm tot}$. Note that
$\chi^2_{\rm Higgs}$ includes only the Higgs-signal strength data while
$\chi^2_{\rm total}$ sums over all experimental datasets:
Higgs+$\left({\cal A}^b_{\rm FB}\right)^{\rm EXP}$+$R^{\rm EXP}_b$+$\Gamma_{\rm had}$.
}
\begin{ruledtabular}
\begin{tabular}{c|ccccc}
Cases & {\bf SM} & {\bf i} & {\bf ii} & {\bf iii} & {\bf iv}  \\
\hline
data    &  & Higgs & $\left({\cal A}^b_{\rm FB}\right)^{\rm EXP}$
      & $R^{\rm EXP}_b$ & $\Gamma_{\rm had}$  
       \\
\hline
\hline
$x_w$        & 0.23154   & 0.23154 & 0.23154 & 0.23154 & 0.23154  \\
$\delta\equiv \Delta/M$  & $0.0$  & $0.253^{+0.063}_{-0.090}$
                         & $0.202^{+0.036}_{-0.046}$ 
                         & $0.0814^{+0.044}_{\rm -limit}$
                         & $0.143^{+0.036}_{-0.052}$ \\
\hline
$C_b\equiv g_{hbb}/g^{\rm SM}_{hbb}$      
             & 1.000   & $0.936^{+0.037}_{-0.036}$  
                       & $0.959^{+0.017}_{-0.016}$
                       & $0.9934^{\rm +limit}_{-0.0091}$ 
                       & $0.980^{+0.012}_{-0.012}$\\
$\chi^2_{\rm Higgs}$    &  53.81     & 50.99  & 51.39 & 53.27  & 52.35  \\
${\cal A}^b_{\rm FB}$ & 0.1030  & 0.0968 & 0.0991 & 0.1024 & 0.1012  \\
$R_b$ & 0.21582  & 0.2208 & 0.2189 & 0.21629 & 0.21731 \\
$\Gamma_{\rm had}$[GeV] & 1.7411 & 1.7523  & 1.7480 & 1.7421 & 1.7444   \\
\hline
$\chi^2_{\rm total}$    &  62.21     & 113.9  & 69.78 & 58.32 & 56.13  \\
\end{tabular}
\begin{tabular}{c|ccc}
Cases & {\bf v} & {\bf Fit-I} & {\bf Fit-II} \\
\hline
      & Higgs+$\left({\cal A}^b_{\rm FB}\right)^{\rm EXP}$ 
& Higgs+$\left({\cal A}^b_{\rm FB}\right)^{\rm EXP}$  & Higgs+
$\left({\cal A}^b_{\rm FB}\right)^{\rm EXP}$  \\
data  & +$R^{\rm EXP}_b+\Gamma_{\rm had}$ 
      &   & +$R^{\rm EXP}_b+\Gamma_{\rm had}$   \\
\hline
\hline
$x_w$           & 0.23154  
                & $0.23109^{+0.00076}_{-0.00082}$ 
                & $0.23202^{+0.00031}_{-0.00031}$  \\
$\delta\equiv \Delta/M$        & $0.132^{+0.022}_{-0.028}$  
                               & $0.253^{+0.063}_{-0.090}$
                               & $0.115^{+0.037}_{-0.027}$ \\
\hline
$C_b\equiv g_{hbb}/g^{\rm SM}_{hbb}$      
             & $0.9826^{+0.0066}_{-0.0063}$ &  $0.936^{+0.037}_{-0.036}$ 
             & $0.9868^{+0.0055}_{-0.0099}$ \\
$\chi^2_{\rm Higgs}$     & 52.53  & 50.99 & 52.80  \\
${\cal A}^b_{\rm FB}$ & 0.10144 & 0.09922 & 0.09918  \\
$R_b$ & 0.21708 & 0.22082 & 0.21677  \\
$\Gamma_{\rm had}$[GeV] & 1.7439 & 1.7523 & 1.7432  \\
\hline
$\chi^2_{\rm total}$     & 55.88  & 113.6 & 53.68   \\
\end{tabular}
\end{ruledtabular}
\end{table}

\begin{figure}[t!]
\centering
\includegraphics[height=2.2in,angle=0]{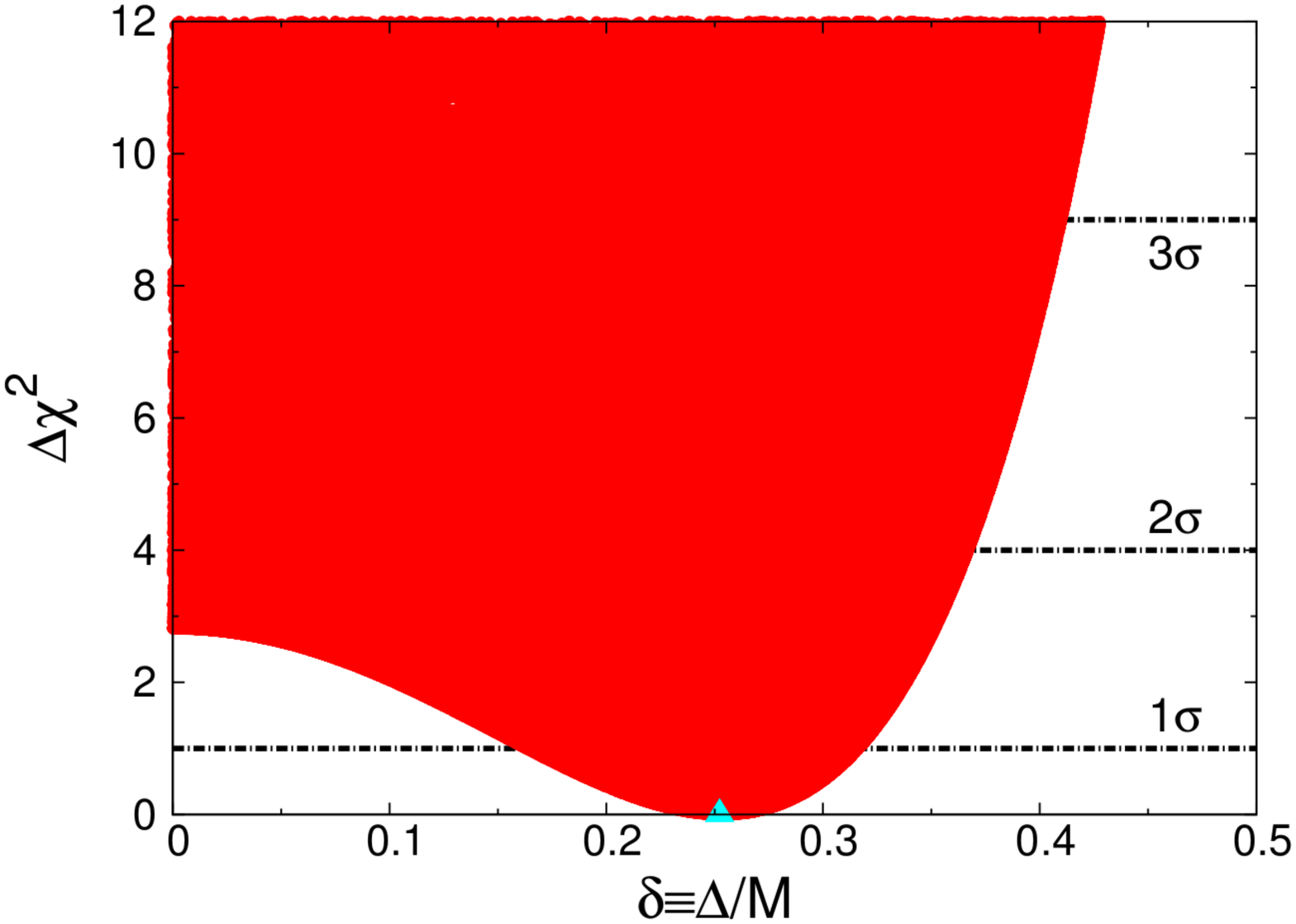}
\includegraphics[height=2.2in,angle=0]{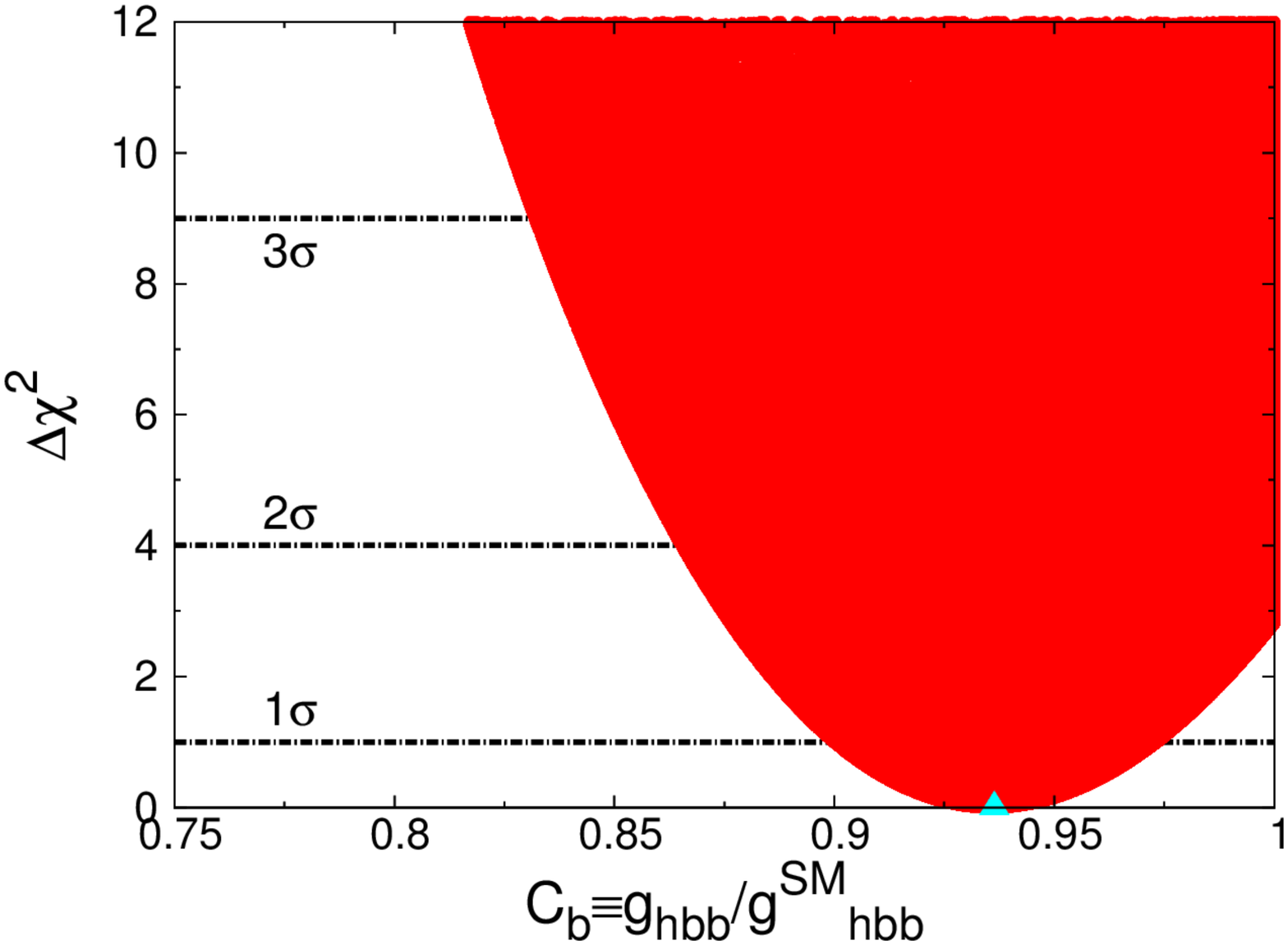}
\includegraphics[height=2.2in,angle=0]{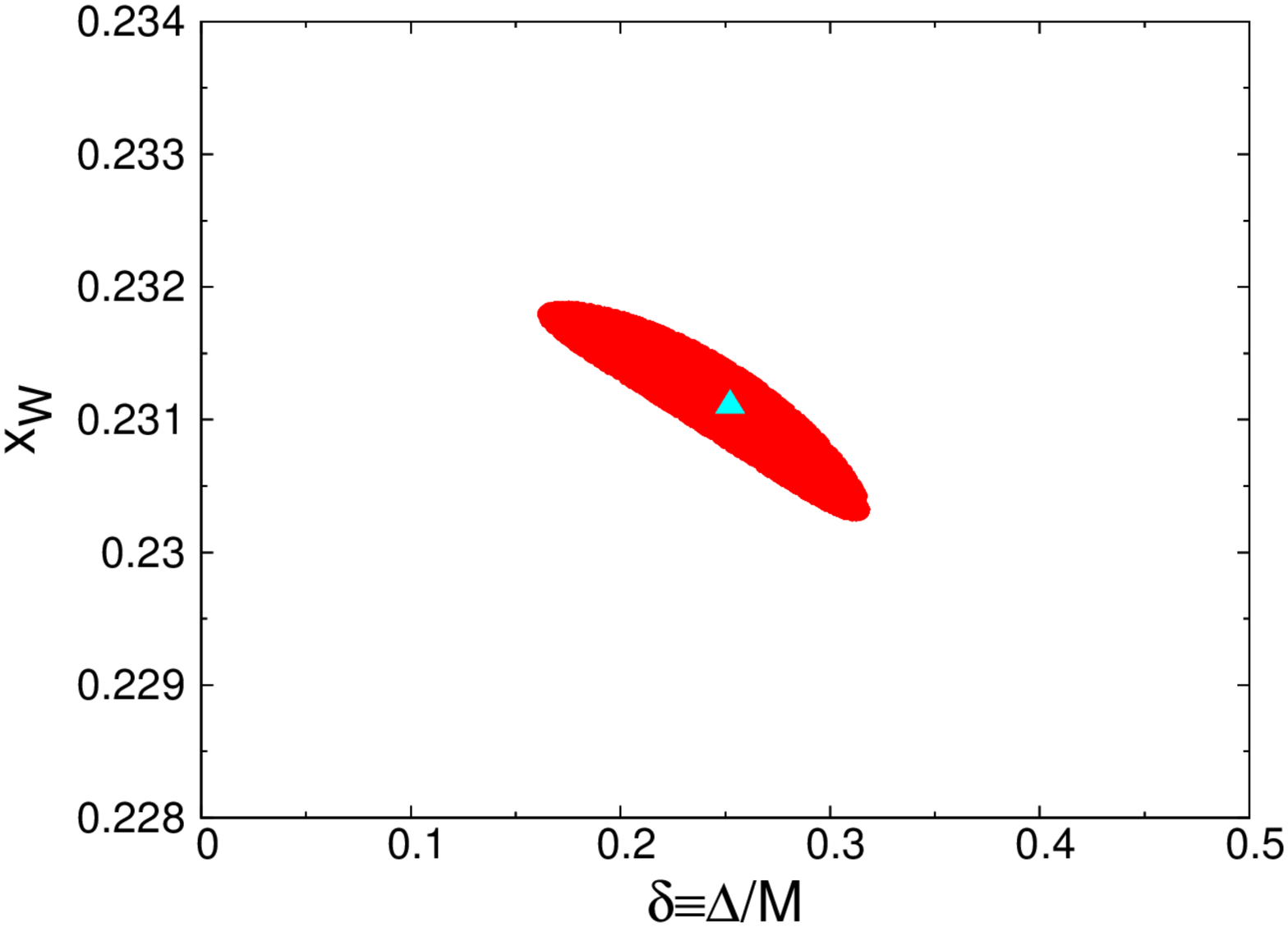}
\includegraphics[height=2.2in,angle=0]{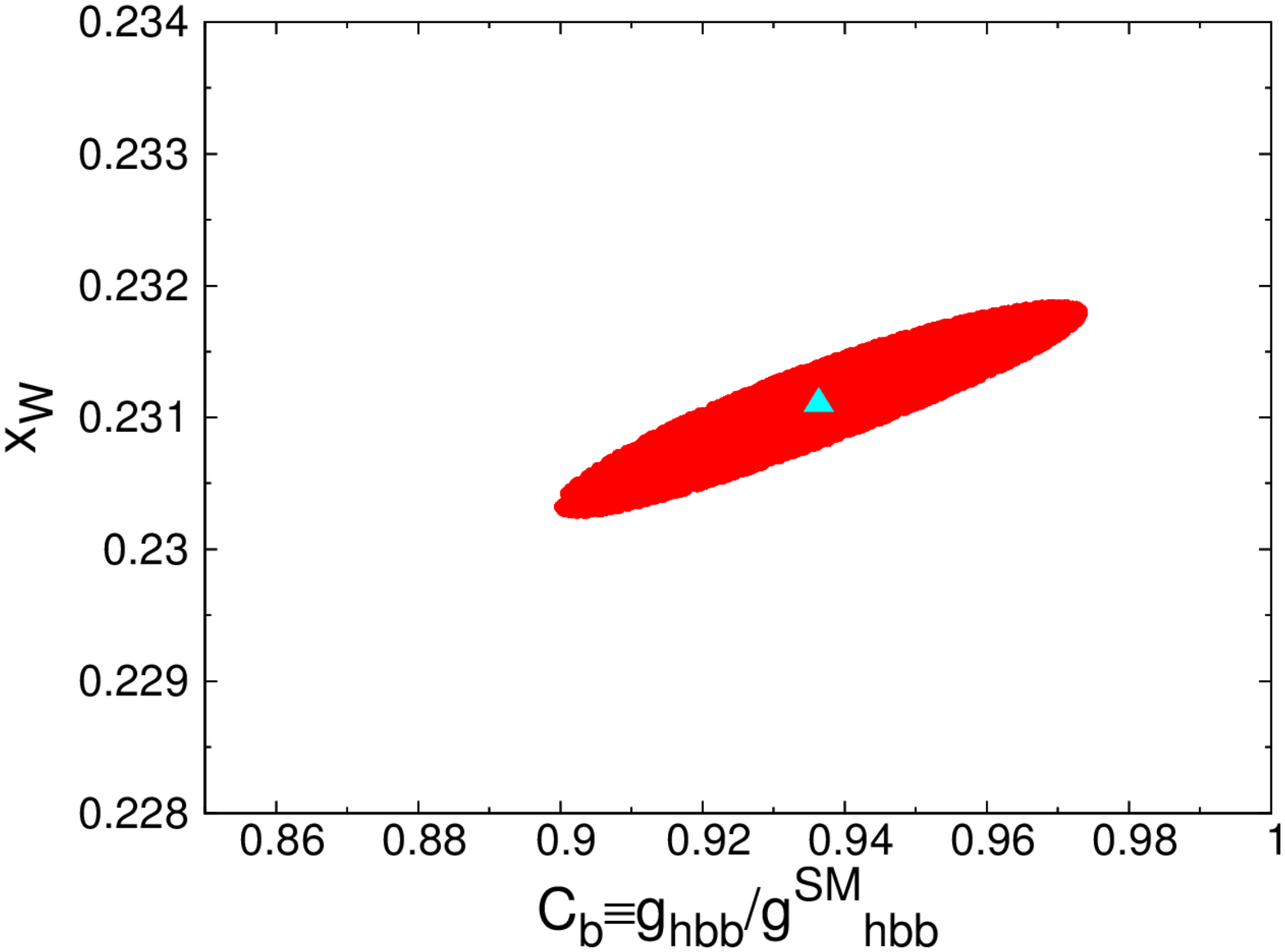}
\caption{\small \label{fig-2} 
{\bf Fit-I:} fitting to the Higgs-signal strengths and 
$({\cal A}^b_{\rm FB})^{\rm EXP}$ datasets by varying $(\delta,x_W)$. 
Upper-panels: $\Delta \chi^2$ distributions versus
$\delta \equiv \Delta /M$ and versus $C_b \equiv g_{h b b} / g_{h b b}^{\rm SM}$.
Lower-panels: the parameter-space region with $\Delta \chi^2 \leq 1$. 
}
\end{figure}

\begin{figure}[t!]
\centering
\includegraphics[height=2.2in,angle=0]{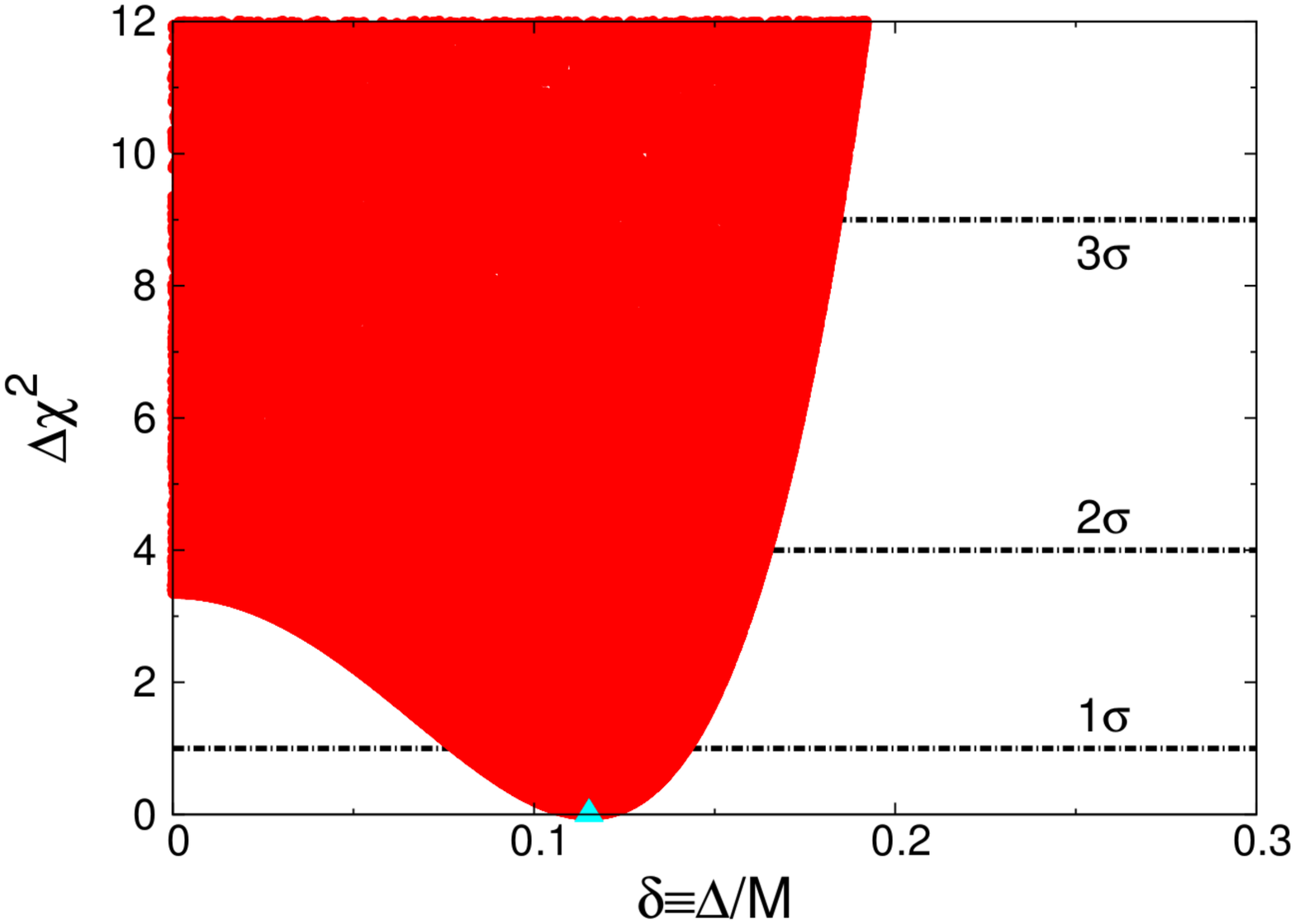}
\includegraphics[height=2.2in,angle=0]{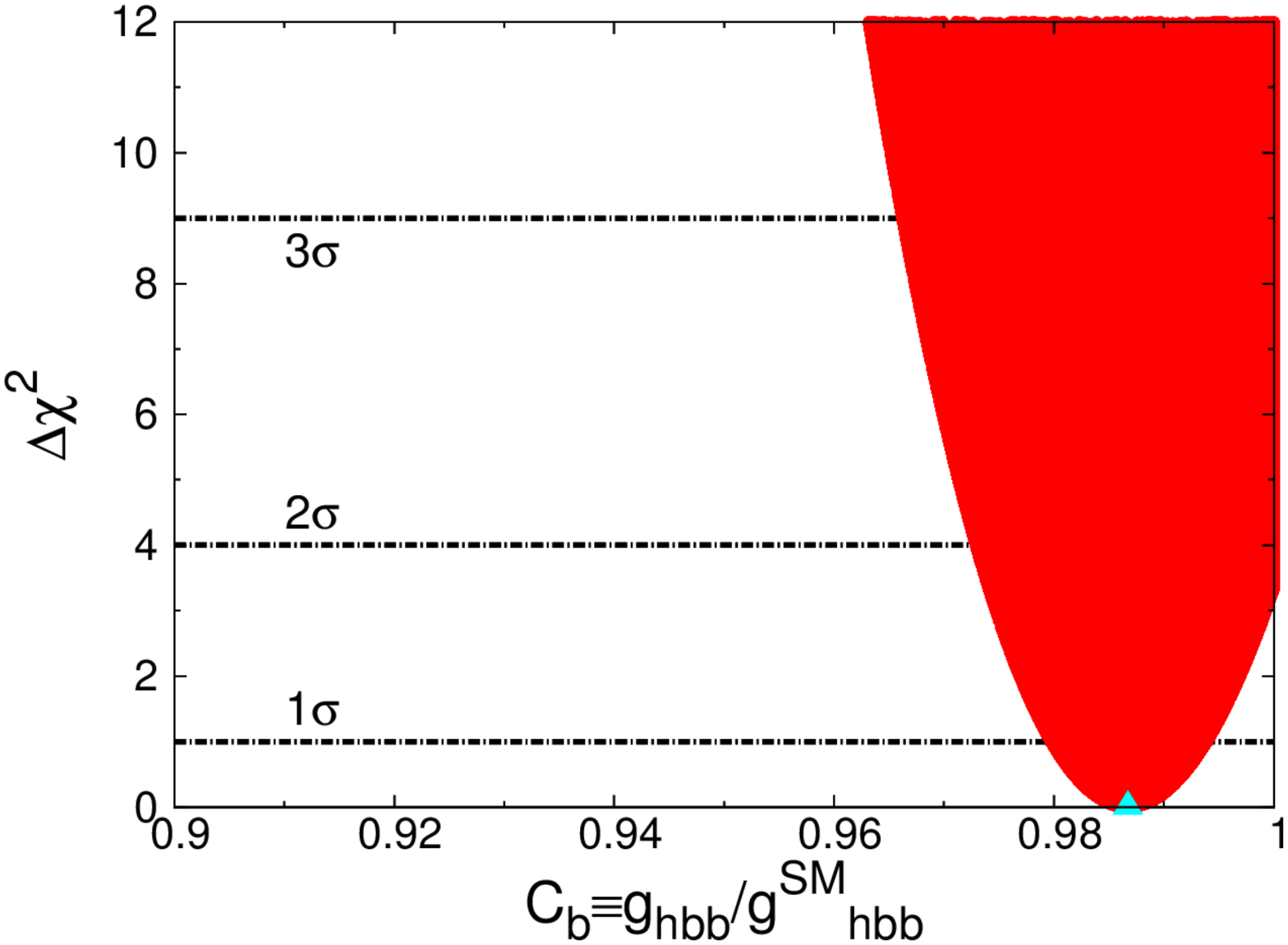}
\includegraphics[height=2.2in,angle=0]{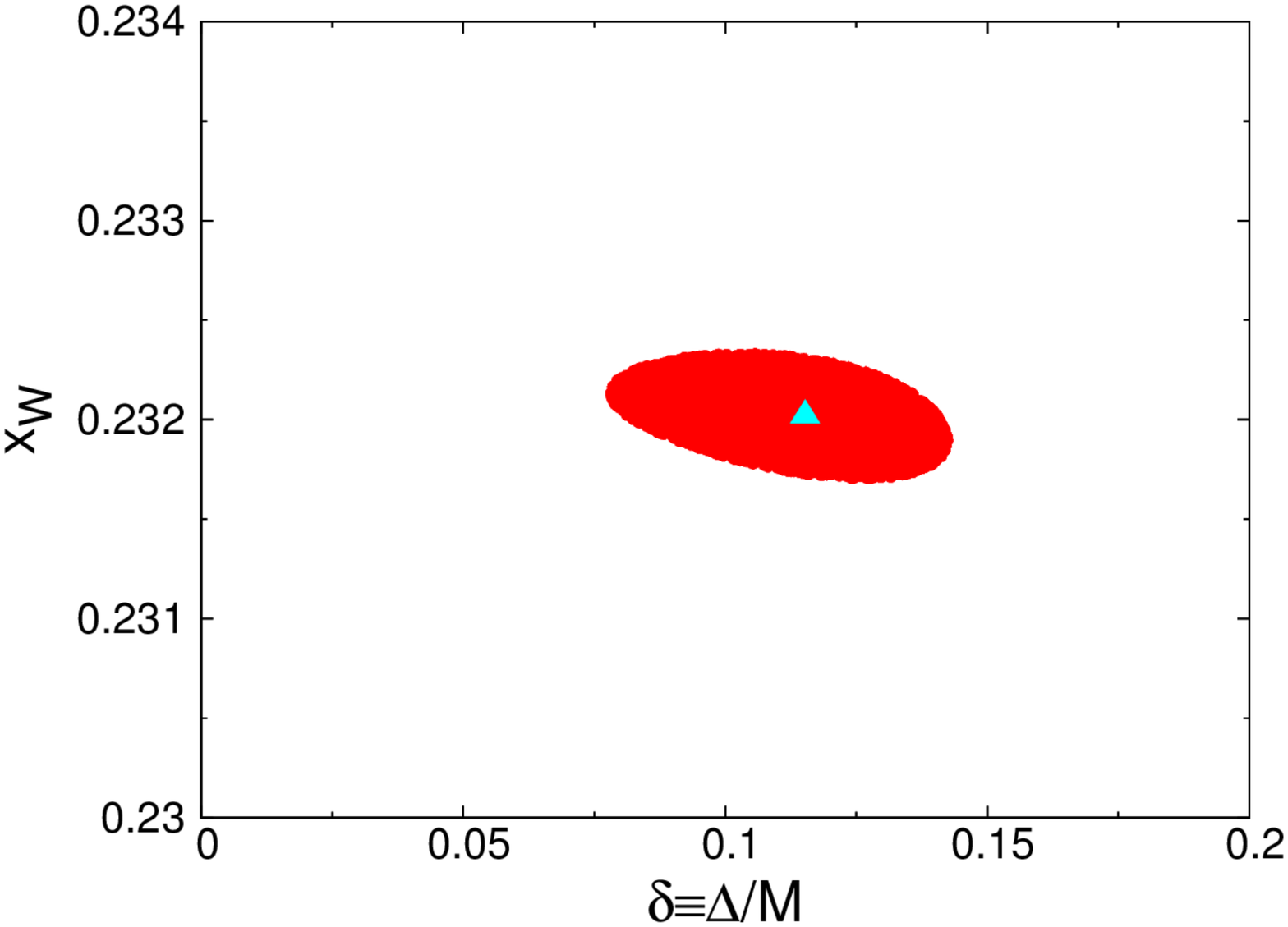}
\includegraphics[height=2.2in,angle=0]{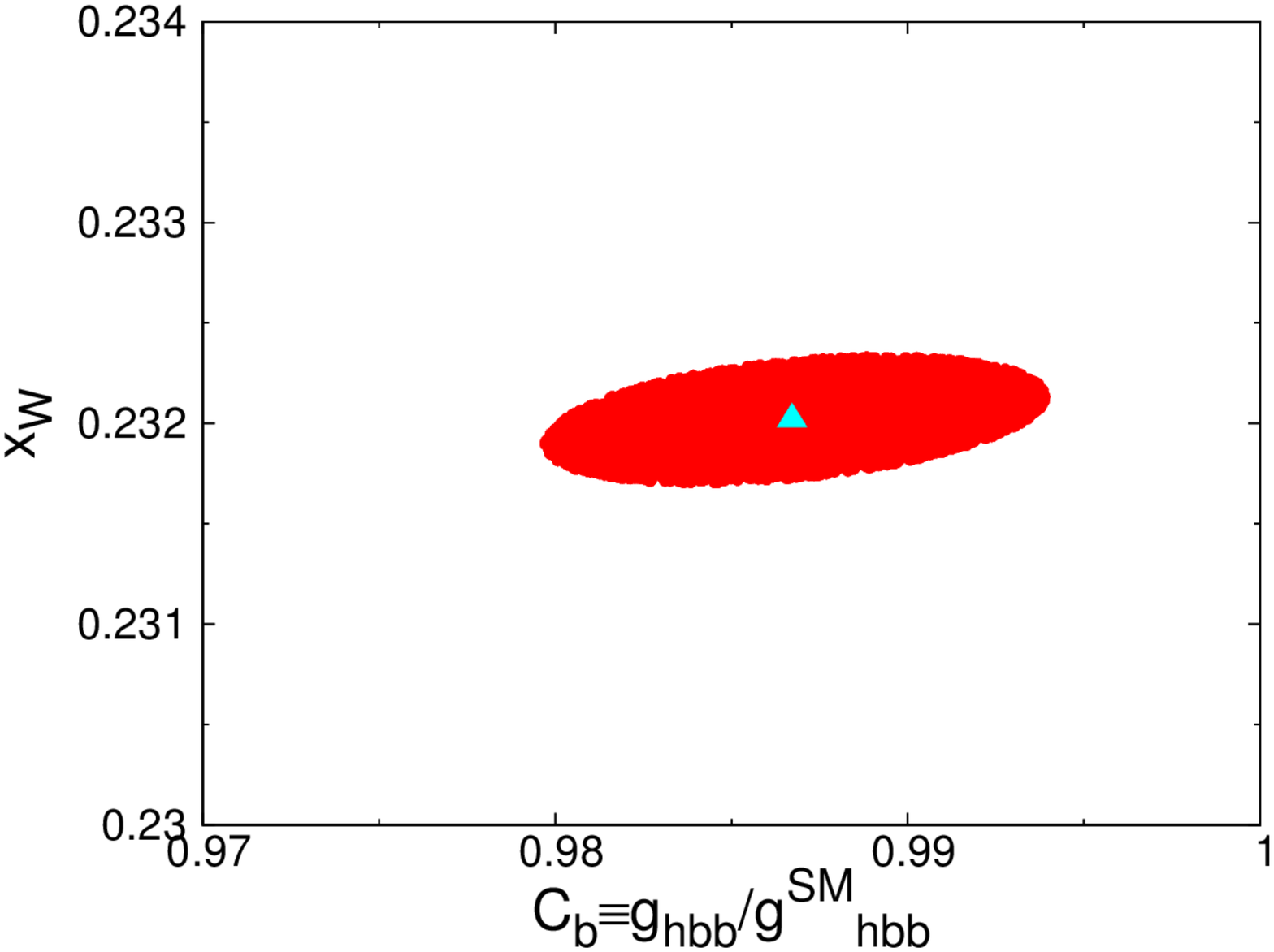}
\caption{\small \label{fig-3} 
{\bf Fit-II:} the same as Fig.~\ref{fig-2}, but fitting to the Higgs-signal strengths,
$({\cal A}^b_{\rm FB})^{\rm EXP}$, $R^{\rm EXP}_b$, and $\Gamma_{\rm had}$ datasets.
}
\end{figure}

  In order to see whether such deviations from SM are
      robust or not, 
we allow the value of $x_w$ floating together with $\delta$, 
and perform two fittings, {\bf Fit-I} and {\bf Fit-II}.
The {\bf Fit-I} only includes the Higgs-signal strengths and 
$\left({\cal A}^b_{\rm FB}\right)^{\rm EXP}$, 
because these two datasets would allow a significant deviation from
the SM, according to the Cases-{\bf i} and {\bf ii}.
The best fit point and $\Delta \chi^2$ distribution are shown in
Table~\ref{tab-2} and Fig.\ref{fig-2}.
In this case, the best fit point
$(\delta, x_w) = (0.25, 0.231)$ gives very good description to the
Higgs-signal strengths and ${\cal A}_{\rm FB}^{\rm EXP}$, but draws a
large deviation in $R_b$ and $\Gamma_{\rm had}$.
For the {\bf Fit-II}, which includes all four datasets,
the best fit values
and $\Delta \chi^2$ distributions are
shown in Table~\ref{tab-2} and Fig.~\ref{fig-3}, respectively.
  The best fit point $(\delta, x_w) = (0.115, 0.232)$
  provides the best description for all four datasets - the lowest
  $\chi^2$ overall.


So far we observe that the Higgs-signal  strengths can be improved
substantially by reducing the bottom Yukawa coupling, which is achieved
in this work by mixing
the RH component $b'_R$ of a vector-like quark doublet
with the SM right-handed bottom quark. So is
the forward-backward asymmetry of the bottom quark at the $Z$ pole.
A mixing of order $s_R \simeq \delta \simeq 0.20 -0.25$ can achieve the effects.
However, such a mixing would deviate $R_b$ and $\Gamma_{\rm had}$ too much.
Overall, a mixing of order
$s_R \simeq \delta \simeq 0.12 -0.13$ would improve the whole picture.

\section{Discussion}

Note that the left-handed $b$ quark mixing is extremely small of order
$m_b v /M^2 \sim 10^{-4}$. All the $B$ decays, including lifetime,
branching ratios, $B^0$-$\overline{B^0}$ mixing and angular distributions,
would not be affected. So 
are the CKM matrix elements, because all these processes
involve the left-handed coupling only. 

The parameter $\delta = \Delta / M = g_{\rm B} v / ( \sqrt{2}  M )
= 0.1 - 0.2$ in the above analysis.  Assuming $g_{\rm B} \sim O(1)$
the mass of the heavy vector-like quark (VLQ) would be of order
$\Delta^2 + M^2 \sim 1- 2 $ TeV.  This VLQ is phenomenologically very
interesting.  It can be directly produced via QCD production processes, such
as $gg, q\bar q \to b' \overline{b'}$ (here $b'$
is understood to be the mass eigenstate).
Assuming the mixing in the left-hand $b'$ is negligible compared to the
right-handed one. the dominant decays of $b'$ are
\[
b' \to b h, \qquad {\rm and} \qquad  b'\to b Z
\]
with partial widths given by
\begin{eqnarray}
  \Gamma(b' \to b h) &=& \left( \frac{\Delta}{v} \right)^2 \frac{M_{b'}}{32\pi}\,
  c_R^2
  \left( 1 - \frac{m_h^2}{M_{b'}^2} \right )^2 \\
  \Gamma(b' \to b Z )&=& \frac{g_Z^2 c_R^2 s_R^2}{128 \pi} \frac{M_{b'}^3}{m_Z^2}
  \, \left ( 1 + \frac{2 m_Z^2}{M_{b'}^2} \right )\,
  \left ( 1 - \frac{m_Z^2}{M_{b'}^2} \right )^2 \,.
\end{eqnarray}
It is understood that the mass of $b'$ is approximately
$\sqrt{\Delta^2 + M^2}$ in the leading
order.  Note that $s_R \approx \Delta /M$ and $c_R \approx 1$ in the limit
$\Delta / M \rightarrow 0$.  The partial width of $b' \to b Z$ can then
be further simplified to
\[
\Gamma(b' \to b Z )= \left( \frac{\Delta}{v} \right )^2 \frac{M_{b'}}{32\pi}
\, \left ( 1 + \frac{2 m_Z^2}{M_{b'}^2} \right )\,
\left ( 1 - \frac{m_Z^2}{M_{b'}^2} \right )^2
\]
Therefore, in the limit $M_{b'} \gg m_Z, m_h$ these two partial widths are
the same. We recall the equivalence theorem that in high energy limit the
Higgs boson and longitudinal mode of gauge bosons behave the same.

The collider signature of pair
production of $b' \overline{b'}$ via
the decay into the $Z$ boson is rather clean
\[
b' \overline{b'} \to (b X)  (\bar b Z) \to (b X) ( \bar b \ell^+ \ell^-)
\]
Such a search for charged lepton pair(s)  plus jets
has been performed at the 13 TeV LHC \cite{Aaboud:2018saj}.
Here  we perform a rough  estimate of the the lower mass limit of $b'$.
The number of events with at least one charged lepton pair is
\begin{equation}
  N = \sigma(pp\to b' \overline{b'}) \times{\cal L} \times
  \left( 1 - B^2(b' \to b h) \right)
  \times B(Z\to \ell^+ \ell^- ) \times \epsilon
\end{equation}
where $\epsilon$ denotes the relevant experimental efficiency collectively.
Taking ${\cal L}=36.1$ fb$^{-1}$, $B(b'\to bh)=0.5$,
$B(Z\to e^+ e^- +\mu^+ \mu^-)= 0.067$, and $\epsilon=0.5$, and
requiring
$N < 4$, we obtain 
\begin{equation}
  \sigma(pp\to b' \overline{b'} ) \lsim 4~{\rm fb} \;.
\end{equation}
This upper limit on production cross section can be translated to
the lower mass limit of $M_{b'} \gsim 1.4$ TeV \cite{Aaboud:2018saj}.

Further searches of $b' \overline{b'} \to (b Z ) (\bar b Z),\;
(b h ) (\bar b Z),\; (b h ) (\bar b h)$ are possible. The signatures
would give 1 or 2 charged lepton pairs at the $Z$ mass plus multiple $b$ jets.

\section*{Acknowledgment}  
W.-Y. K. and P.-Y. T. thank the National Center of Theoretical Sciences, 
Taiwan, R.O.C. for hospitality. 
The work of K.C. was supported by the National Science
Council of Taiwan under Grants Nos. MOST-105-2112-M-007-028-MY3 and
MOST-107-2112-M-007-029-MY3.
The work of J.S.L. was supported by
the National Research Foundation of Korea (NRF) grant
No. NRF-2016R1E1A1A01943297.
The work of P.-Y.T. was supported by World Premier International Research 
Center Initiative (WPI), MEXT, Japan. 

\bibliography{reference}
\bibliographystyle{h-physrev}

\end{document}